\documentclass{svmult}
\usepackage[english]{babel}
\usepackage{amsmath,amssymb,amscd,graphicx,bbm}

\usepackage{makeidx}         
\usepackage{graphicx}        
\usepackage{multicol}        
\usepackage[bottom]{footmisc}

\makeindex             

\catcode`\>=\active \def>{
\fontencoding{T1}\selectfont\symbol{62}\fontencoding{\encodingdefault}}
\catcode`\|=\active \def|{
\fontencoding{T1}\selectfont\symbol{124}\fontencoding{\encodingdefault}}
\newcommand{\mathd}{\mathrm{d}}
\newcommand{\mathe}{\mathrm{e}}
\newcommand{\nocomma}{}
\newcommand{\nosymbol}{}
\newcommand{\nobracket}{}

\newcommand{\tmop}[1]{\ensuremath{\operatorname{#1}}}

\begin{document}

\title*{Frontiers of open quantum system dynamics}
\author{Bassano Vacchini}
\institute{Dipartimento di Fisica, Universit\`a degli Studi di Milano
\\
Via Celoria 16, Milan, 20133, Italy\\
INFN Sezione di Milano\\ Via Celoria 16, Milan, 20133, Italy
\\
\texttt{bassano.vacchini@mi.infn.it}}
%
%
\maketitle

\begin{abstract}
  We briefly examine recent developments in the field of open quantum system
  theory, devoted to the introduction of a satisfactory notion of memory for a
  quantum dynamics. In particular, we will consider a possible formalization
  of the notion of non-Markovian dynamics, as well as the construction of
  quantum evolution equations featuring a memory kernel. Connections will be
  draw to the corresponding notions in the framework of classical stochastic
  processes, thus pointing to the key differences between a quantum and
  classical formalization of the notion of memory effects.
\end{abstract}

\section{Introduction}
The theory of open quantum systems denotes the application of quantum
mechanics to situations in which the dynamics of the quantum system of
interest is influenced by other degrees of freedom, that we are neither
interested nor capable to take into account in detail. While in principle any
system has to be considered open, since a perfect shielding from the
environment is never feasible, in simple situations isolation can be
considered as a good approximation. In many realistic settings however, the
effect of an external quantum environment on the system dynamics cannot be
neglected. This is the case e.g. in many instances of quantum optics,
condensed matter physics and quantum chemistry. For the case of an open system
dynamics, many interesting physical effects and mathematical structures do
appear {\cite{Breuer2002}}. From the physical viewpoint, with respect to a
closed dynamics we are faced with new phenomena like dissipation and
decoherence, which only have partial analog in the classical setting. Such
phenomena play a crucial role in many relevant recent fields of research, such
as quantum computation and quantum thermodynamics
{\cite{Nielsen2000,Millen2016a,Alicki2018a}}.

In this research field many problems are still open, which have important
connections to mathematics. In this contribution we will try to highlight, in
a concise way, some recent developments in the field of open quantum systems,
connected to some of the presently most active research lines,
relevant for the mathematical formulation of the theory. After a brief
description of the framework of open quantum systems in Sect.~2, we will
address in Sect.~3 the delicate question of the definition of non-Markovian
quantum processes. This implies introducing a notion of non-Markovian
dynamics, which is quite different form the classical one, though the two
can be naturally connected. In Sect.~4 we will point to the derivation of
equations of motions allowing to introduce memory effects, focusing in
particular on master equations with a memory kernel, for which again a natural
connection to a class of non-Markovian classical processes can be considered.

\section{Open quantum system dynamics}

\begin{figure}[h]
\label{figura} 
  \begin{center}
    \includegraphics{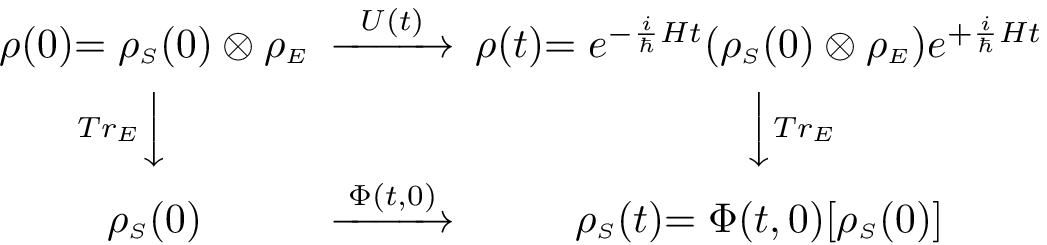}
  \end{center}
  \caption{Commutative diagram showing the existence of a reduced dynamics for
  an initial system-environment state in factorized from. The reduced state of
  the system at time $t$ can be equivalently obtained by taking the marginal
  with respect to the environmental degrees of freedom of the unitarily
  evolved total state, or by applying the completely positive trace preserving
  map $\Phi ( t,0 )$ to the initial state of the system.}
\end{figure}

Let us first introduce some basic elements of open quantum system theory
{\cite{Breuer2002}}, which actually consists in considering the dynamics of a
quantum system described on the Hilbert space $\mathcal{H}_{S}$ without
assuming it to be isolated, so that it interacts with an external environment
described on a Hilbert space $\mathcal{H}_{E}$ by menas of unitary operators
$U ( t )$ acting on $\mathcal{H}_{S} \otimes \mathcal{H}_{E}$. The fact that
the system is not closed brings with itself two important new aspects. On the
one hand, the reduced dynamics of the isolated system only is not described by
a Liouville von-Neumann equation, and purity of the state is not preserved
during the evolution, On the other hand, even a factorized system-environment
state develops correlations, so that the latter play a major role in the time
evolution. The tensor product structure of the underlying Hilbert space, on
its turn, brings in two important aspects. On the one hand, states can exhibit
correlations which are of non-classical nature, such as entanglement. On the
other hand, the evolution of the reduced system as a function of time is
described by a collection of transformations which have the property of being
completely positive, a property strictly connected to the non-commutativity of
the space of observables. Assuming that the state at the initial time is
factorized
\begin{eqnarray}
  \rho_{SE} ( 0 ) & = & \rho_{S} ( 0 ) \otimes \rho_{E} ,  \label{eq:prod}
\end{eqnarray}
we have that the reduced state of the system at a later time is given by
\begin{eqnarray}
  \rho_{S} ( t ) & = & \tmop{Tr}_{E} \{ U ( t ) \rho_{S} ( 0 ) \otimes
  \rho_{E} U ( t )^{\dag} \} , \label{eq:rd} 
\end{eqnarray}
where $\tmop{Tr}_{E}$ denotes the partial trace with respect to the
environmental degrees of freedom. This state contains all the information
relevant for the description of the dynamics of the system observables. In
particular this transformation defines a linear map
\begin{eqnarray}
  \Phi ( t,0 ) [ \rho_{S} ( 0 ) ] & \equiv & \tmop{Tr}_{E} \{ U ( t ) \rho_{S}
  ( 0 ) \otimes \rho_{E} U ( t )^{\dag} \} ,  \label{eq:fit}
\end{eqnarray}
which considering an orthogonal resolution for the state of the environment
$\rho_{E} = \sum_{\xi} \lambda_{\xi} P_{\varphi_{\xi}}$, and introducing an
orthogonal basis $\{ \varphi_{\eta} \}$ in $\mathcal{H}_{E}$, admits the
following representation
\begin{eqnarray}
  \Phi ( t,0 ) [ \rho_{S} ( 0 ) ] & = & \sum_{\xi , \eta} \lambda_{\xi}
  \langle \varphi_{\eta} | U ( t ) \nobracket \varphi_{\xi} \rangle \rho_{S} (
  0 ) ( \langle \varphi_{\eta} | U ( t ) \nobracket \varphi_{\xi} \rangle
  )^{\dag} \nonumber\\
  & = & \sum_{\xi , \eta} K_{\xi \eta} ( t ) \rho_{S} ( 0 ) K_{\xi
  \eta}^{\dag} ( t ) ,  \label{eq:Kraus}
\end{eqnarray}
where we have introduced so-called Kraus operators $K_{\xi \eta} =
\sqrt{\lambda_{\xi}} \langle \varphi_{\eta} | U ( t ) \nobracket \varphi_{\xi}
\rangle$ acting on $\mathcal{H}_{S}$.
This representation warrants complete positivity of the
map. A map $\Phi$ defined on the space of trace class operators $\mathcal{T} (
\mathcal{H}_{S} )$ is said to be completely positive if its extension $\Phi
\otimes \mathbbm{1}_{n}$ to $\mathcal{T} ( \mathcal{H}_{S} \otimes
\mathbbm{C}^{n} )$ defined on operators in tensor product form as
\begin{eqnarray}
  \Phi \otimes \mathbbm{1}_{n} [ A \otimes B ] & = & \Phi [ A ] \otimes B
  \nonumber
\end{eqnarray}
is a positive map for any $n\in \mathbbm{N}$. Otherwise stated, the trivial extension of a
completely positive map acting on some system, to a larger set of degrees of
freedom the system is not interacting with, remains positive. It can be shown
that any such map admits the representation Eq.~(\ref{eq:Kraus}), and
viceversa {\cite{Kraus1983}}. For an initial state in factorized form as in
Eq.~(\ref{eq:prod}) it is thus possible to define a reduced dynamics,
described by the time dependent collection of completely positive trace
preserving maps $\Phi ( t,0 )$ given by Eq.~(\ref{eq:fit}), as shown
in Fig.~(\ref{figura}).
Two natural questions appear at this stage. On the one hand, now that
reversibility of the dynamics warranted by the unitary evolution has got lost,
it is interesting to ascertain whether such maps do describe memory effects. On
the other hand, one would like to know the possible expression of maps $\Phi (
t,0 )$ describing a well defined dynamics, as well as the general structure of
evolution equations for the statistical operator admitting such collection of
maps as solution. These two aspects have been the object of extensive
research, and some recent developments in this respect will be discussed
Sects. 3 and 4.

\section{Characterization of dynamics with memory}

The existence of the reduced dynamics for an open quantum system implies that
its time evolution can be described by evolution equations which, on top of a
coherent quantum dynamics as can be obtained by a Liouville-von Neumann
equation, do exhibit stochasticity. The stochastic contribution to the
dynamics is due to the interaction with the unobserved quantum degrees of
freedom of the environment. A quite natural question in this setting, in
analogy with what happens for a classical stochastic dynamics, is therefore
whether the obtained quantum dynamics can exhibit effects which can be
reasonably termed memory effects. In the description of classical systems
random features in the dynamics are described by the mathematically well
established notion of stochastic process. The notion of lack of memory for a
stochastic process is enforced by asking a suitable constraint on the
conditional probabilities determining the process. Roughly speaking, a process
is defined to be Markovian, that is without memory, if the only relevant
conditioning of the probability densities for the outcomes of the considered
stochastic process is with respect to the last ascertained value of the time
dependent random variable considered, and not with respect to values at
previous times. In such a way a notion of memory is naturally introduced (see
e.g. {\cite{Cox1965}} for a proper formalization of this notion). For a Markov
process the notion of lack of memory is therefore naturally linked to the
neglecting of knowledge of values taken by the random variable in the past.

In the quantum framework, random variables have to be described by
self-adjoint operators acting in the Hilbert space $\mathcal{H}_{S}$ of the
considered system. However, in order to obtain the probability distribution
for the values taken by such random variables at a given time one has to
perform a measurement. At variance with the classical case, knowledge of the
value of the random variable will thus affect the subsequent evolution of the
system in a non negligible way, depending on the way the measurement is
performed. The external intervention necessary for the measurement thus
influences the value of multitime probability densities, which do not admit an
obvious definition as in the classical case. The definition of a quantum
Markovian process along the lines of the classical viewpoint, pursued in the
first systematic studies on the characterisation of open quantum system dynamics
{\cite{Lewis1981a,Accardi1982a,Lindblad1979a}}, thus encounters major
difficulties.

More recently, different approaches have been considered, which tackle the
issue considering features of the dynamics determined by quantities depending
on a single point in time, rather than on multitime probability densities (see
{\cite{Breuer2012a,Rivas2014a,Breuer2016a,Devega2017a}} for reviews). In such
a way one can overcome the difficulties related to the measurement problem and
allow for a direct experimental verification of the property of Markovianity.
The connection between these approaches and the notion of classical Markovian
process has been discussed in {\cite{Vacchini2011a}}. In this contribution we
will concentrate on a approach which connects non-Markovianity to a reversible
exchange of information between system and environment, started with the
seminal paper {\cite{Breuer2009b}}. Let us consider a reduced dynamics on
$\mathcal{H}_{S}$ defined by a collection of completely positive trace
preserving maps $\Phi ( t,0 )$ sending the initial system state to the state
at time $t$
\begin{eqnarray}
  \Phi & \equiv & \{ \Phi ( t,0 ) \}_{\nocomma t \in \mathbbm{R}_{+}} ,
  \nonumber
\end{eqnarray}
and suppose that the experimenter can prepare two distinct initial states of
the system, say $\rho_{S}^{1} ( 0 )$ and $\rho_{S}^{2} ( 0 )$, with the same
probability $p=1/2$. If an observer has to guess which state has actually been
prepared by performing a single measurement, as shown in {\cite{Helstrom1976}}
the maximal probability of success, obtained by performing an optimal
measurement, is given by the expression
\begin{eqnarray}
  P ( 0 ) & = & \frac{1}{2} ( 1+D ( \rho_{S}^{1} ( 0 ) , \rho_{S}^{2} ( 0 ) )
  ) ,  \label{eq:po}
\end{eqnarray}
where
\begin{eqnarray}
  D ( \rho , \sigma ) & = & \frac{1}{2} \| \rho - \sigma \|_{1} \nonumber
\end{eqnarray}
denotes the trace distance between two statistical operators $\rho , \sigma
\in \mathcal{T} ( \mathcal{H}_{S} )$, namely the normalized distance built by
means of the trace norm $\| \cdot \|_{1}$. If the observer tries to
distinguish the states at a later time $t$, after interaction of the system
with the environment, the success probability is now given by
\begin{eqnarray}
  P ( t ) & = & \frac{1}{2} ( 1+D ( \rho_{S}^{1} ( t ) , \rho_{S}^{2} ( t ) )
  ) ,  \label{eq:pt}
\end{eqnarray}
where $\rho_{S}^{1,2} ( t ) = \Phi ( t,0 ) [ \rho_{S}^{1,2} ( 0 ) ]$ and due
to a crucial property of the trace distance we have
\begin{eqnarray}
  P ( t ) & \leqslant & P ( 0 ) . \nonumber
\end{eqnarray}
Indeed the trace distance is a contraction under the action of an arbitrary
positive, and therefore in particular completely positive, trace preserving
transformation $\Lambda$ {\cite{Ruskai1994a}}
\begin{eqnarray}
  D ( \Lambda [ \rho ] , \Lambda [ \sigma ] ) & \leqslant & D ( \rho , \sigma
  ) . \nonumber
\end{eqnarray}
The effect of the interaction with the environment is thus a reduction of the
capability to distinguish quantum states of the system by an observer
performing measurements on the system only. For the case in which the dynamics
is characterised by a semigroup composition law so that
\begin{eqnarray}
  \Phi ( t,0 ) & = & \mathe^{t \mathcal{L}} ,  \label{eq:qds}
\end{eqnarray}
with $\mathcal{L}$ a suitable generator, corresponding to dynamics typically
called quantum Markov processes, one further has
\begin{eqnarray}
  P ( t ) & \leqslant & P ( s ) \hspace{2em} \forall t \geqslant s, \nonumber
\end{eqnarray}
so that there is a monotonic decrease in time of the distinguishability
between system states. This feature is taken as the defining property of a
quantum Markovian dynamics. Accordingly, a quantum dynamics described by a
collection of completely positive trace preserving maps $\Phi ( t,0 )$ is said
to be non-Markovian if there are revivals in time in the success probability
$P ( t )$ or equivalently in the trace distance $D ( \rho_{S}^{1} ( t ) ,
\rho_{S}^{2} ( t ) )$, which contains the relevant part of the information, so
that
\begin{eqnarray}
  \dot{D}  ( \rho_{S}^{1} ( t ) , \rho_{S}^{2} ( t ) ) & \geqslant & 0, 
  \label{eq:nm}
\end{eqnarray}
for at least a point in time and a couple of initial states, where we have
denoted by $\dot{D}$ the time derivative of the trace distance between the
evolved initial system states.

These revivals do generally depend on the choice of initial states, so that a
suitable quantifier of non-Markovianity of the dynamics has been introduced
according to the expression
\begin{eqnarray}
  \mathcal{N} ( \Phi ) & = & \max_{\rho^{1,2}_{S} ( 0 )}   \int_{\dot{D} >0}
  dt \hspace{0.27em} \dot{D}  ( \rho_{S}^{1} ( t ) , \rho_{S}^{2} ( t ) ) . 
  \label{eq:N}
\end{eqnarray}
It immediately appears that this definition of non-Markovian quantum dynamics
only requires to observe the state of the system at different times and
starting from different system initial conditions, rather than on multitime
quantities, so that an experimental assessment of non-Markovianity can be
obtained by means of a tomographic procedure {\cite{Liu2011a}}.

To substantiate the interpretation of this notion of non-Markovianity as
information back flow from the environment to the system, let us introduce the
following quantities {\cite{Nielsen2000,Breuer2016a}}
\begin{eqnarray}
  \mathcal{I}_{\mathrm{int}} ( t ) & = & D ( \rho_{S}^{1} (t), \rho_{S}^{2}
  (t))  \label{eq:int}
\end{eqnarray}
and
\begin{eqnarray}
  \mathcal{I}_{\mathrm{ext}} ( t ) & = & D ( \rho_{SE}^{1} (t), \rho_{SE}^{2}
  (t)) -D ( \rho_{S}^{1} (t), \rho_{S}^{2} (t)) ,  \label{eq:ext}
\end{eqnarray}
where $\mathcal{I}_{\mathrm{int}} ( t )$ is used to quantify the internal
information, that is the information accessible by performing measurements on
the system only, while $\mathcal{I}_{\mathrm{ext}} ( t )$ denotes the external
information, which can only be obtained by performing measurements in the
Hilbert space of both system and environment $\mathcal{H}_{S} \otimes
\mathcal{H}_{E}$, minus the internal one. If the overall dynamics is unitary
the sum of the two quantities $\mathcal{I}_{\mathrm{tot}} ( t ) =
\mathcal{I}_{\mathrm{int}} ( t ) + \mathcal{I}_{\mathrm{ext}} ( t )$ is
conserved, so that in particular
\begin{eqnarray}
  \frac{\mathd}{\tmop{dt}} D ( \rho_{S}^{1} (t), \rho_{S}^{2} (t)) & = &
  \frac{\mathd}{\tmop{dt}} \mathcal{I}_{\mathrm{int}} ( t ) \nonumber\\
  & = & - \frac{\mathd}{\tmop{dt}} \mathcal{I}_{\mathrm{ext}} ( t ) \nosymbol
  . \nonumber
\end{eqnarray}
This equality shows that an increase in time of the trace distance corresponds
to a decrease in the external information, which being overall conserved can
only flow from the environment into the system. To understand in which sense
information can be stored outside the system, namely it cannot be retrieved by
performing measurements on the system only, it is enlightening to consider the
following bound, first introduced in Ref.~{\cite{Laine2010b}} in connection to
detection of initial correlations
\begin{eqnarray}
  D ( \rho^{1}_{S} ( t ) , \rho^{2}_{S} ( t ) ) -D ( \rho^{1}_{S} ( s ) ,
  \rho^{2}_{S} ( s ) ) & \leqslant & D ( \rho^{1}_{SE} ( s ) , \rho^{1}_{S} (
  s ) \otimes \rho^{1}_{E} ( s ) )  \label{eq:bound}\\
  &  & +D ( \rho^{2}_{SE} ( s ) , \rho^{2}_{S} ( s ) \otimes \rho^{2}_{E} ( s
  ) ) \nonumber\\
  &  & +D ( \rho^{1}_{E} ( s ) , \rho^{2}_{E} ( s ) ) , \nonumber
\end{eqnarray}
where it is assumed that $t \geqslant s$ and, at variance with
{\cite{Laine2010b}}, $\rho^{1,2}_{SE} ( 0 ) = \rho^{1,2}_{S} ( 0 ) \otimes
\rho_{E} ( 0 )$, where $\rho^{1,2}_{S} ( 0 )$ and $\rho_{E} ( 0 )$ are the marginal states
obtained by taking the partial trace with respect to the degrees of freedom of
environment and system respectively, so as to ensure the existence of a reduced dynamics. As
discussed above the trace distance can be naturally understood as a quantifier
of distinguishability among quantum states, so that in particular, if the two
statistical operators are a state on a bipartite space and the product of its
marginals as in the r.h.s. of Eq.~(\ref{eq:bound}), it provides a quantifier
of correlations in the overall state. The l.h.s. corresponds to the change
over time in trace distance, which can only be positive if at least one of the
quantities at the r.h.s. is different from zero, that is after interacting for
a time $s$ either system and environment have become correlated or the
environmental state has changed in different ways depending on the initial
system state.

This definition of non-Markovianity of a quantum dynamics based on the notion
of information back flow between system and environment is strictly connected
to an alternative notion relying on a mathematical property of the collection
of completely positive trace preserving maps describing the reduced dynamics.
Indeed such a collection is called $P$-divisible if the following identity
holds {\cite{Rivas2012}}
\begin{eqnarray}
  \Phi ( t,0 ) & = & \Phi ( t,s ) \Phi ( s,0 ) \hspace{2em} \forall t
  \geqslant s \geqslant 0,  \label{eq:div}
\end{eqnarray}
with $\Phi ( t,s )$ positive maps for any $t \geqslant s \geqslant 0$, while
it is called $CP$-divisible if the maps $\Phi ( t,s )$ are in particular
completely positive for any $t \geqslant s \geqslant 0$, as in the case e.g.
of the quantum dynamical semigroup considered in Eq.~(\ref{eq:qds}). It
immediately appears that both $CP$-divisible and $P$-divisible are Markovian
according to the trace distance criterion defined above, while a monotonic
decrease of the trace distance in general does not warrant neither kind of
divisibility. The composition law Eq.~(\ref{eq:div}) tells us that, in order
to predict the time evolution of the system forward in time, we only need to
know the state at a given time, thus naturally inducing a formalization of
lack of memory, and indeed its violation was proposed as a definition of
non-Markovian dynamics in {\cite{Rivas2010a}}. \

\subsection{Generalized non-Markovianity measure}

Recently different refinements of the definition and quantification of
non-Markovianity as given by formulae Eq.~(\ref{eq:nm}) and Eq.~(\ref{eq:N})
have been considered {\cite{Wissmann2012a,Wissmann2015a,Breuer2018a}},
importantly always supporting the seminal interpretation of non-Markovianity
as information back flow from environment to system. An important and natural
generalization, first suggested in {\cite{Chruscinski2011a}}, consists in
considering the discrimination problem between quantum states, used to connect
trace distance and distinguishability, in the more general setting in which
the two known states $\rho_{S}^{1} ( 0 )$ and $\rho_{S}^{2} ( 0 )$ can be
prepared with different weights, say $p_{1}$ and $p_{2}$. In this case the
optimal strategy can be shown to lead to the following success probability
\begin{eqnarray}
  P ( 0 ) & = & \frac{1}{2} ( 1+ \Delta ( \rho_{S}^{1} ( 0 ) , \rho_{S}^{2} (
  0 ) ;p_{1} ,p_{2} ) ) ,  \label{eq:poHel}
\end{eqnarray}
where the expression $\Delta ( \rho_{S}^{1} ( 0 ) , \rho_{S}^{2} ( 0 ) ;p_{1}
,p_{2} ) = \| p_{1} \rho_{S}^{1} ( 0 ) -p_{2} \rho_{S}^{2} ( 0 ) \|$ is also
known as norm of the Helstrom matrix. Non-Markovianity is then identified with
a revival in time of the norm of the Helstrom matrix
\begin{eqnarray}
  \dot{\Delta}  ( \rho_{S}^{1} ( t ) , \rho_{S}^{2} ( t ) ;p_{1} ,p_{2} ) &
  \geqslant & 0,  \label{eq:nmHelm}
\end{eqnarray}
for at least a point in time, a couple of initial states and a choice of
weights, which provide apriori information on the prepared state. Note that
the class of processes which are non-Markovian is thus enlarged, including
situations which where previously not encompassed
{\cite{Wang2013a,Wissmann2015a}}. Accordingly, a generalized measure of
non-Markovianity can be considered, given by the expression
\begin{eqnarray}
  \mathcal{N} ( \Phi ) & = & \max_{p_{1,2} , \rho^{1,2}_{S} ( 0 )}  
  \int_{\dot{\Delta} >0} dt \hspace{0.27em} \dot{\Delta} ( \rho_{S}^{1} ( t )
  , \rho_{S}^{2} ( t ) ;p_{1} ,p_{2} ) .  \label{eq:NHelm}
\end{eqnarray}
An important result of this generalization of the initial definition is the
fact that it allows for a clearcut connection with the notion of divisibility
considered in Eq.~(\ref{eq:div}), which in its mathematical formulation does
not immediately show a link to information flow, since the latter can
only be formulated by introducing a quantifier of distinguishability among
states. Indeed, thanks to a result by Kossakowski connecting the positivity
property of a trace preserving map with its contractivity when acting on an
arbitrary hermitian observable {\cite{Kossakowski1972b}}, monotonicity in time
of the behavior of the norm of the Helstrom matrix $\Delta ( \rho_{S}^{1} ( t
) , \rho_{S}^{2} ( t ) ;p_{1} ,p_{2} )$ can be shown to be equivalent to
$P$-divisibility of the collections of time evolution maps in the sense of
Eq.~(\ref{eq:div}), provided the time evolution map is invertible as a linear
map on the space of operators. Most importantly, this extension is still
compatible with the notion of information back flow as characterizing a
non-Markovian dynamics. This fact can be shown considering suitable
generalizations of the notion of internal and external information as
considered in Eq.~(\ref{eq:int}) and Eq.~(\ref{eq:ext}), as well as a
generalisation of the bound Eq.~(\ref{eq:bound}), thus pointing to the general
validity of the starting definition of quantum non-Markovianity
{\cite{Breuer2018a,Amato2018a}}.

\section{Non-Markovian evolution equations}

As discussed in Sect.~2, considering an initially factorized state of system
and environment is sufficient to warrant the existence of a reduced dynamics,
which will depend both on the state of the environment and the unitary
interaction, according to expression Eq.~(\ref{eq:rd}). However, in the
general case the evaluation of the exact dynamics is utterly unfeasible, so
that it is of utmost importance to have access to approximate methods. On the
one hand, one can consider perturbation expansions; on the other hand, one can
look for phenomenological expressions. In both cases one major difficulty is
warranting that the obtained time evolutions indeed provide a well-defined
dynamics, corresponding to a completely positive trace preserving
transformation. In particular, the requirement of complete positivity, which
warrants connection to an underlying microscopic dynamics, is difficult to be
enforced and is typically lost at intermediate steps in a perturbative
approach. A fundamental result has been obtained for the situation in which
the time evolution, instead of obeying a group evolution law as in the case of
a reversible unitary dynamics, can be described by a semigroup, thus
introducing a preferred direction in time. In this case the collection of maps
is called quantum dynamical semigroup and is determined by a generator
$\mathcal{L}$ according to Eq.~(\ref{eq:qds}). A fundamental theorem of open
quantum system theory [14, 15] states that this generator has to be in the
so-called Gorini-Kossakowksi-Sudarshan-Lindblad form
\begin{eqnarray}
  \mathcal{L} [ \rho_{S} ] & = & - \frac{i}{\hbar} [ H, \rho_{S} ] + \sum_{k}
  \gamma_{k} \left[ L_{k} \rho_{S} L_{k}^{\dag} - \frac{1}{2} \{ L_{k}^{\dag}
  L_{k} , \rho_{S} \} \right] ,  \label{eq:lind}
\end{eqnarray}
with $H$ a self-adjoint operator corresponding to an effective Hamiltonian,
$\gamma_{k}$ positive rates and $L_{k}$ system operators also called Lindblad
operators. It provides a generalization of the Liouville von-Neumann equation
to include both decoherence and dissipative effects. Solutions of the time
evolution equation
\begin{eqnarray}
  \frac{\mathd}{\tmop{dt}} \rho_{S} ( t ) & = & \mathcal{L} [ \rho_{S} ( t ) ]
  \label{eq:l}
\end{eqnarray}
together with a suitable initial condition $\rho_{S} ( 0 )$ do define a
collection of completely positive trace preserving maps obeying a semigroup
composition law. As discussed in Sect.~3, the obtained dynamics is Markovian
and provides the quantum analog of a classical semigroup evolution. To
describe memory effects more general dynamics have to be considered. To this
aim one can either consider time dependent generalizations of the generator
considered in Eq.~(\ref{eq:lind}), or move to evolution equations explicitly
featuring a memory kernel. In both cases one has to ensure that the solutions
of such equations do provide a collection of time dependent completely
positive trace preserving maps, thus describing a well defined dynamics. In
the case of so-called time local evolution equations, one has to replace rates
and operators appearing in Eq.~(\ref{eq:lind}) by time dependent quantities,
looking for conditions warranting complete positivity. Considering master
equations in integrodifferential form
\begin{eqnarray}
  \frac{\mathd}{\tmop{dt}} \rho_{S} ( t ) & = & \int_{0}^{t} \mathd \tau
  \mathcal{K} ( t- \tau ) [ \rho_{S} ( \tau ) ]  \label{eq:k}
\end{eqnarray}
the corresponding task is to envisage conditions on the operator kernel
$\mathcal{K} ( t )$ warranting preservation of positivity and trace of the
solutions of the integrodifferential equation. In both cases the most general
solution to the problem is not known, even not heuristically, while partial
results have been recently obtained
{\cite{Budini2004a,Chruscinski2014a,Kossakowski2009a,Vacchini2013a,Vacchini2014a,Chruscinski2016a,Vacchini2016b,Chruscinski2016a,Chruscinski2017a}}.
In particular we will consider how to obtain well-defined quantum memory
kernels $\mathcal{K} ( t )$. While quantum dynamical semigroups can be seen as
the quantum counterpart of classical Markov semigroups, the class of
considered memory kernels can be taken as the quantum analogue of a class of
non-Markovian processes known as semi-Markov process {\cite{Nollau1980}}.

We consider as starting point an expression for the exact solution of
Eq.~(\ref{eq:lind}), which can be written as
\begin{eqnarray}
  \rho_{S} (t) & = & \Phi ( t,0 ) [ \rho_{S} ( 0 ) ]  \label{eq:path}\\
  & = & \mathcal{R} ( t ) [ \rho_{S} ( 0 ) ] \nonumber\\
  &  & + \sum_{k=1}^{\infty} \int_{0}^{t} \hspace{-0.17em} \hspace{-0.17em}
  \tmop{dt}_{k} \ldots \hspace{0.27em} \int_{0}^{t_{2}} \hspace{-0.17em}
  \hspace{-0.17em} \tmop{dt}_{1} \mathcal{R} ( t-t_{k} ) \mathcal{J} \ldots
  \mathcal{R} ( t_{2} -t_{1} ) \mathcal{J} \mathcal{R} ( t_{1} ) [ \rho_{S} (
  0 ) ] , \nonumber
\end{eqnarray}
where we have introduced the contraction semigroup
\begin{eqnarray}
  \mathcal{R} ( t ) [ \nosymbol \rho ] & = & \mathe^{- \frac{i}{\hbar} Ht-
  \frac{1}{2} \sum_{_{k}} \gamma_{k} L_{k}^{\dag} L_{k} t} \rho \mathe^{+
  \frac{i}{\hbar} Ht- \frac{1}{2} \sum_{_{k}} \gamma_{k} L_{k}^{\dag} L_{k} t}
  \nonumber
\end{eqnarray}
and the completely positive map
\begin{eqnarray}
  \mathcal{J} [ \rho ] & = & \sum_{k} \gamma_{k} L_{k} \rho L_{k}^{\dag} .
  \nonumber
\end{eqnarray}
As a result the solution is expressed as a sum of contributions characterised
by a given number of insertions of the completely positive map $\mathcal{J}$
with an intermediate trace decreasing evolution in between. Complete
positivity of the overall evolution is warranted by the fact that we are
considering sum and composition of completely positive maps, which form a
convex cone. The solution is expressed in a space of ``trajectories''
determined by the number of jumps or insertions of the map $\mathcal{J}$ and
the points in time at which these jumps happen
{\cite{Barchielli2009,Holevo2001}}. Both maps $\mathcal{R} ( t )$ and
$\mathcal{J}$ are determined by the rates $\gamma_{k}$ and the Lindblad
operators $L_{k}$. To consider a more general situation one can define a
collection of linear maps in analogy with Eq.~(\ref{eq:path}), introducing the
replacement of jump operator and contraction semigroup by means of an
arbitrary completely positive trace preserving transformation $\mathcal{E}$
and a collection of time dependent completely positive trace preserving maps
$\mathcal{F} ( t )$, according to the scheme
\begin{eqnarray}
  \rho_{S} (t) & = & g ( t ) \mathcal{F} ( t ) [ \rho_{S} ( 0 ) ] 
  \label{eq:gen}\\
  &  & + \sum_{k=1}^{\infty} \int_{0}^{t} \hspace{-0.17em} \hspace{-0.17em}
  \tmop{dt}_{k} \ldots \hspace{0.27em} \int_{0}^{t_{2}} \hspace{-0.17em}
  \hspace{-0.17em} \tmop{dt}_{1} f ( t-t_{k} ) \mathcal{F} ( t-t_{k} )
  \mathcal{E} \ldots \nonumber\\
  &  & \hspace{2em} \times \ldots f ( t_{2} -t_{1} ) \mathcal{F} ( t_{2}
  -t_{1} ) \mathcal{E} g ( t_{1} ) \mathcal{F} ( t_{1} ) [ \rho_{S} ( 0 ) ] .
  \nonumber
\end{eqnarray}
In the representation Eq.~(\ref{eq:gen}) we have further inserted the
functions $f ( t )$ and $g ( t )$. The function $f ( t )$ has to be positive
and normalized to one over the interval $[ 0, \infty )$, so as to be
interpreted as a waiting time distribution. Accordingly, the function $g ( t
)$ is determined by $\dot{g} ( t ) =-f ( t )$, together with $g ( 0 ) =1$, so
that it can be interpreted as the associated survival probability. It can be
easily seen that these properties are sufficient to identify the linear
assignment $\rho_{S} ( 0 ) \rightarrow \rho_{S} ( t )$ obtained through
Eq.~(\ref{eq:gen}) as a collection of completely positive trace preserving
maps. These evolutions correspond to a situation in which in between the
evolution given by the maps $\mathcal{F} ( t_{} )$, the system undergoes a
transformation described by the completely positive trace preserving map
$\mathcal{J}$. It is however not obvious the existence of a closed evolution
equation for the statistical operator of the system $\rho_{S} (t)$, so as to
connect the transformations to a continuous dynamics. To this aim one
considers the expression of the Laplace transform of Eq.~(\ref{eq:gen}), which
thanks to the presence of convolutions takes the simple form
\begin{eqnarray}
  \widehat{\rho_{}}_{S} ( u ) & = & ( \mathbbm{1} - \widehat{f \mathcal{F}} (
  u ) \mathcal{E} )^{-1} \widehat{g \mathcal{F}} ( u ) \rho_{S} ( 0 ) ,
  \nonumber
\end{eqnarray}
where the hat denotes the Laplace transform, so that by a suitable
rearrangement one has
\begin{eqnarray}
  u \hat{\rho}_{S} ( u ) - \rho_{S} ( 0 ) & = & \left[ \frac{1}{\widehat{g
  \mathcal{F}} ( u )} \widehat{f \mathcal{F}} ( u ) \mathcal{E} - \left(
  \frac{1}{\widehat{g \mathcal{F}} ( u )} -u \right) \right] \hat{\rho}_{S} (
  u ) ,  \label{eq:cksmp}
\end{eqnarray}
allowing to identify the memory kernel $\mathcal{K} ( t )$ in Eq.~(\ref{eq:k})
with the inverse Laplace transform of the operator
\begin{eqnarray}
  \widehat{\mathcal{K}} ( u ) & = & \frac{1}{\widehat{g \mathcal{F}} ( u )}
  \widehat{f \mathcal{F}} ( u ) \mathcal{E} - \left( \frac{1}{\widehat{g
  \mathcal{F}} ( u )} -u \right) ,  \label{eq:kk}
\end{eqnarray}
thus showing in particular that indeed the transformation Eq.~(\ref{eq:gen})
describes a closed dynamics. Actually it can be shown that the kernel
Eq.~(\ref{eq:kk}) despite its complex expression does have a simple and
natural interpretation and allows for a connection with a class of
non-Markovian processes known as semi-Markov {\cite{Breuer2008a,Breuer2009a}}.
These generally non-Markovian classical processes describe a dynamics in a
discrete state space, in which jumps from site $m$ to site $n$ take place with
probability given by the elements of a stochastic matrix $\pi_{nm}$, at times
distributed according to the waiting time distribution $f_{n} ( t )$. For
these processes one can introduce a generalized master equation obeyed by the
one-point probability density $P_{n} ( t )$ given by
{\cite{Feller1964a,Nollau1980}}
\begin{eqnarray}
  \frac{d}{dt} P_{n} (t) & = & \int_{0}^{t} d \tau \sum_{m} [ W_{nm} ( \tau
  )P_{m} (t- \tau )-W_{mn} ( \tau )P_{n} (t- \tau ) ] , \nonumber
\end{eqnarray}
whose expression in Laplace transform reads
\begin{eqnarray}
  u \hat{P}_{n} (u) -P_{n} (0) & = & \sum_{m} \left[ \pi_{nm}
  \frac{\hat{f}_{m} (u)}{\hat{g}_{m} (u)} - \delta_{nm} \left(
  \frac{1}{\hat{g}_{m} (u)} -u \right) \right]  \hat{P}_{m}  (u) . 
  \label{eq:clu}
\end{eqnarray}
A natural correspondence can be drawn between Eq.~(\ref{eq:clu}) and
Eq.~(\ref{eq:cksmp}). The stochastic matrix $\pi_{nm}$ is replaced by the
completely positive trace preserving map $\mathcal{E}$, while the collection
of waiting time distributions $f_{n} ( t )$ goes over to $f ( t ) \mathcal{F}
( t_{} )$, product of waiting time distribution and completely positive trace
preserving maps. Classical functions are therefore now replaced by operators.
The classical dynamics corresponding to jumps between sites with probabilities
determined by a given stochastic transition matrix and at times dictated by
given waiting time distributions, is replaced by a piecewise quantum dynamics
in the space of statistical operators. In this quantum dynamics
transformations described by a completely positive trace preserving
map $\mathcal{E}$, at
times described by a fixed waiting time distribution, are interspersed with a
continuous time evolution described by the collection of completely positive
trace preserving maps $\mathcal{F} ( t )$. It immediately appears that in the
correspondence from Eq.~(\ref{eq:clu}) to Eq.~(\ref{eq:cksmp}) an important
and typically quantum feature appears, namely the relevance of operator
ordering. Indeed Eq.~(\ref{eq:cksmp}) can have different quantum counterparts,
and another operator ordering leads to an alternative expression for the
kernel
\begin{eqnarray}
  \widehat{\mathcal{K}} ( u ) & = & \mathcal{E} \widehat{f \mathcal{F}} ( u )
  \frac{1}{\widehat{g \mathcal{F}} ( u )} - \left( \frac{1}{\widehat{g
  \mathcal{F}} ( u )} -u \right) ,  \label{eq:kkl}
\end{eqnarray}
which substituted in Eq.~(\ref{eq:k}) still leads to a well-defined dynamics.
Indeed it turns out that the two combinations describe different microscopic
modelling of a quantum piecewise dynamics. The microscopic dynamics formalised
by Eq.~(\ref{eq:kk}) corresponds to the physics of the micromaser
{\cite{Raithel1994a,Cresser1992a,Cresser1996a}}, while the kernel
Eq.~(\ref{eq:kkl}) naturally appears in so-called collision models
{\cite{Lorenzo2016a,Lorenzo2017a}}.

\section{Conclusions and outlook}

We have briefly exposed recent work within the framework
of open quantum system theory, aiming at the definition and quantification of
the so-called non-Markovianity, to be understood as the capability of a
quantum dynamics to feature memory effects. In particular, we have pointed to
a notion of non-Markovian dynamics connected to an information exchange
between the considered system and the surrounding environment, whose
generalization can be naturally connected to a notion of divisibility of
quantum maps. We have further considered a possible extension of a known class
of master equations describing a completely positive trace preserving dynamics
to include memory effects by means of the introduction of a memory kernel.

Great efforts are presently being put in the endeavour to understand the
relevance of the proposed notions of non-Markovian quantum dynamics for the
description of relevant physical systems (see in this respect the recent reviews
{\cite{Breuer2012a,Rivas2014a,Breuer2016a,Devega2017a}}). A critical and
important open issue is, in particular, whether it captures distinctive
features of the dynamics, or if a non-Markovian evolution brings with itself
advantages in performing relevant tasks, e.g. in quantum information or
quantum thermodynamics.

\section*{Acknowledgements}

The author acknowledges support from the EU Collaborative Project QuProCS
(Grant Agreement 641277) and from MIUR through the FFABR project.


\end{document}